# Isotopic Resonance Hypothesis: Experimental Verification by *Escherichia coli* Growth Measurements

*Xueshu Xie[1] and Roman A. Zubarev[1]\**

1. Division of Physiological Chemistry I, Department of Medical Biochemistry and Biophysics, Karolinska Institutet, Stockholm, Sweden

\*Corresponding author: Roman.Zubarev@ki.se   phone/fax +46 8 524 87 594

**Abstract**

Isotopic composition of reactants affects the rates of chemical and biochemical reactions. As a rule, enrichment of heavy stable isotopes leads to slower reactions. But the recent isotopic resonance hypothesis suggests that the dependence of the reaction rate upon the enrichment degree is not monotonous; instead, at some "resonance" isotopic compositions, the kinetics increases, while at "off-resonance" compositions the same reactions progress slower. To test the predictions of this hypothesis for the elements C, H, N and O, we designed a precise (standard error ±0.05%) experiment to measure the bacterial growth parameters in minimal media with varying isotopic compositions. A number of predicted resonance conditions were tested, which kinetic enhancements as strong as +3% discovered at these conditions. The combined evidence extremely strongly supports the existence of isotopic resonances. This phenomenon has numerous implications for the origin of life and astrobiology, and possible applications in agriculture, biotechnology, medicine and other areas.

**Introduction**

The four elements C, H, O and N (CHON) have fundamental importance. They are among the seven most abundant elements in our galaxy[1]. Together with helium, CHON are the five most common elements in the Solar system[2]. In Earth's crust, oxygen is the most abundant element, while C, H and N are among the dozen of most abundant elements[3]. In atmosphere, CHON are among the five most abundant elements. In humans, CHON account for 96% of the body weight[4]. In bacteria, CHON compose 92% of the dry mass, and over 98% of the total living mass. CHON are the dominant elements in biopolymers, such as proteins, nucleic acids, polysaccharides and lipids. All four elements have several stable isotopes, with the lightest isotope dominating (e.g., $^2$H or deuterium contributes to the mass of hydrogen 0.0156%, or 156 ppm).

Immediately after the discovery by Urey *et al.* in 1932 of deuterium[5], the biological effects of this heavy isotope have been intensively studied. It has been quickly found that highly enriched deuterium oxide ("heavy water") negatively affects growth and well-being of many organisms. Large excess of deuterium in water was found to reduce protein and nucleic acids synthesis, disturb cell division and alter cellular morphology[6]. High amounts of deuterium were proven toxic to higher organisms[7], although bacteria are able to adapt to grow in almost pure heavy water[8]. There have been much fewer reports on the effect of other heavy stable isotopes in biology. In general, only high enrichments have produced significant alterations. Katz *et al.* grew *Chlorella vulgaris* at high enrichment of deuterium together with other heavy isotopes ($^{13}$C, $^{15}$N, and $^{18}$O) and have reported multiple abnormal biological effects[9]. But later it was found possible to grow mice, sometimes for several generations, in the environment highly enriched with $^{13}$C[10], $^{18}$O[11], and $^{15}$N[12]. The heavy isotopes of C, N and O are currently considered "safe"[11], although recently Turck *et al.* have reported that mice growth on [15]N diet exhibit systematic behavioural differences[12]. They also found that *E. coli* grow slower in a media highly enriched with [15]N[13].

These examples demonstrate that the effects of heavy isotopic substitution on biology is still insufficiently understood[14]. But the most poorly investigated are the effects of low enrichment levels, e.g., of deuterium.

The biological effects of dilute heavy water have first been reported by a Yale researcher T. C. Barnes. In his experiments, mass cultures of *Spirogyra* exhibited much less abscission or cell disjunction (sign of growing at favourable conditions) and greater longevity in 0.06 % D water compared to ordinary water[15]. The positive effect of low deuterium concentration was further demonstrated in experiments confirming an increased longevity in *Spirogyra*[16]. In flatworms, longevity in the heavy water media appeared at deuterium concentrations of 0.06% and 0.07%, while at 0.13% and up to 0.47% D the effect became progressively obscure[17]. Increased cell division was observed in *Euglena* kept for forty-five days in 0.06 % heavy water[18]. Richards has repeated Barnes' tests with yeast and confirmed that a slight excess of deuterium in water may be biologically significant. He observed a 26% increase in dry weight of yeast grown in 0.06% D water[19]. The strongest effect of 0.06% deuterium was observed during fast cellular proliferation[20]. The flatworm *Phagocatagracilis* left in normal water gradually shrank after a few months time to one-fifth or less of their original body size while in those left in 0.06% D water showed only a slight diminution[17]. Lockemann and Leunig studied the effect of heavy water with ≤0.54% D upon *Escherichia coli* (*E. coli*) and *Pseudomonasa eruginosa*. Concentrations of as low as 0.04% D were found to favour survival at adverse conditions[21].

Not all groups have confirmed the effect of diluted heavy water. Macht and Davis reported no difference between the growth in ordinary water and that with 0.06% D[22]. Curry *et al.*

repeated experiments by Barnes and others but could not confirm the previously reported effects of dilute heavy water[23]. In fact, their data did imply a 10% faster growth, but the result was not statistically significant due to a large experimental uncertainty[23]. In general, early reports on the effects of diluted heavy water can be easily criticized for insufficient statistics and the small number of concentration points (often just two) used in the studies.

In 1970s, Lobyshev *et al.* have realized the deficiencies of previous efforts and studied the effects of low deuterium concentrations on biological and biochemical systems with much greater rigor. They found that the Na,K-ATPase activity increases at low deuterium concentrations, reaching maximum (+50% compared to normal water) at 0.04-0.05% D[24,25]. They also studied regeneration of hydroid polyps *Obelia geniculata* in a wide range of $D_2O$ added to sea water. They found strong inhibition at high deuterium concentrations, as well as activation of regeneration by small (≤0.1%) deuterium concentrations[26].

In 1990s, Somlyai *et al.* have also shown that 0.06% deuterium in tissue culture activated the growth of $L_{929}$ fibroblast cell lines[27]. In contrast, water with deuterium depleted below the normal levels suppressed fast-growing cells[27–29].

In many of the above reports, small deuterium concentrations gave a measureable effect, which was in relative terms greater (and often much greater) than the ratio between the deuterium and hydrogen atoms. Another common feature for these studies was the absence of a convincing explanation for this phenomenon, although Lobyshev *et al.* understood that the effect must be collective in nature[25].

Recently, Zubarev *et al.* have formulated the Isotopic Resonance hypothesis that provides a plausible framework for the above results[30]. The hypothesis predicts that at certain "resonance" abundances of the stable isotopes of C, H, N and O, the rates of chemical and biochemical reactions of certain compound classes accelerate, affecting biological growth. The proposed mechanism relates to the overall reduction of the system's quantum mechanical complexity.

The isotopic resonance conditions become obvious after plotting the normalized isotopic shift (NIS, the difference between the average and monoisotopic molecular masses normalized by the nominal mass, which is an integer number) against the normalized monoisotopic defect (NMD, the difference between the monoisotopic and nominal masses of the molecule normalized by the nominal mass)[31]:

NMD =1000*(Monoisotopic mass – nominal mass)/(nominal mass);

NIS = 1000*(Average isotopic mass – monoisotopic mass)/(nominal mass).

For instance, mapping ca. 3,000 tryptic peptides from *E. coli* on such a 2D plot produces, besides the expected scattered "galaxy", a gap with a line that crosses the "galaxy" (Figure 1a). The gap and the line represent an "isotopic resonance". The line appears due to a specific property of terrestrial isotopic compositions of CHON, while the gap is due to the fact that the peptide molecules consist of discrete number of atoms, which determines the discrete character of molecular masses, and thus of the monoisotopic defects and isotopic shifts.

Isotopic resonances, i.e. straight lines in the 2D mass plot, can be observed at many different sets of isotopic compositions. At such a resonance, the number of independent parameters describing the mass of the molecular system is reduced, which results in reduction of system's complexity. The isotopic resonance hypothesis postulates that this reduction affects (usually accelerates) the rates of chemical and biochemical reactions. If the hypothesis is correct, then the terrestrial resonance in Figure 1a may have aided life emerging and taking root on our planet[30].

The line obtained at standard terrestrial isotopic compositions is not perfect, and can be further "tuned up" to become mathematically perfect. Achieving this can be done by varying the isotopic composition of any member of the CHON family; e.g., by increasing the deuterium content from the normal 0.016% to 0.03-0.06%[32]. At a perfect resonance, the rates of biochemical reactions should increase compared to terrestrial conditions. Thus reaching the perfect resonance can explain the startling effects of ultralow deuterium enrichment[15–27]. On the other hand, further deviation from the perfect resonance, e.g. by depletion of deuterium in water, should decelerate the growth, which may explain the anticancer properties of water with depleted deuterium[28,29]. Interestingly, deep depletion that practically removes deuterium from consideration, decreases the system complexity compared to moderate depletion, and thus the hypothesis predicts that deep depletion should increase the reaction rates once again.

To test the isotopic resonance hypothesis, we have previously analyzed available data from published literature, and found an agreement, sometimes remarkable one, with hypothesis' predictions[32]. Recently, we have designed a very precise (standard error ±0.05%), robotically aided experiment measuring the growth parameters of *E. coli* in M9 minimal media (composed of water, glucose, ammonium chloride and inorganic salts) with varying isotopic compositions. The first study performed with the new set-up concerned the effects of low and ultralow deuterium enrichment[33]. In short, previously reported phenomena at ≈0.03% D have been confirmed. Although the magnitude of the effect at 39 °C was small (<1%), bacteria did appear to grow more comfortably at 0.03-0.06% D than at normal value.

Here we continued testing the isotopic resonance hypothesis on the same set-up but for other, non-terrestrial resonances. As in the deuterium study, we monitored three growth parameters that are measured independently: the lag phase duration, the maximum growth rate and the maximum density of bacteria. More comfortable growth conditions usually result in shorter lag phase, faster growth rate and higher maximum density, even though exceptions related to the last parameter have been found at >1% D[33]. Resonances are predicted for $^{15}$N at ≈3.5% (the standard terrestrial value is 0.37%), or $^{13}$C at ≈0.3% (1.1%), or $^{18}$O at 6.6% (0.2%).

These resonances are predicted to be of different "strengths". One of the strongest possible resonances should be at simultaneous enrichment of $^{13}$C to 9.54%, $^{15}$N to 10.89% and $^{18}$O to 6.6%. This "super-resonance" was investigated in detail.

**Results and Discussion**

*Testing the $^{15}$N ≈3.5% resonance*

Preliminary experiments showed that the growth rate of *E. coli* is significantly retarded at higher content of $^{15}$N, which is in line with literature[13]. For instance, at 50%, the lag time was extended by 0.53%, the growth rate decreased by 0.77%, and the maximum density reduced by 0.94% (Figure S1). Thus the null hypothesis was that at ≈3.5% $^{15}$N enrichment, the growth of *E. coli* will be slightly suppressed: linear extrapolation of the 50% $^{15}$N results expected the lag time extended by 0.04%, the growth rate decreased by 0.05%, and the maximum density reduced by 0.07%. In the experiment, a statistically significant *increase* in the maximum growth rate was observed. Figure 2a summarizes the results of seven independent experiments where the nitrogen isotopic composition was varied from 0.37% (normal) to 10%, each experiment involving 32 sample/standard pairs for each $^{15}$N content point. The maximum effect was found at ≈3% $^{15}$N, where the growth rate increased by ≈0.5% (p = 0.007 in two-tailed Student's test), and it was associated with the largest spread of the data. A similar effect has been observed in experiments with deuterium[33]. Indeed, since the initial bacterial composition was genetically and epigenetically heterogeneous, and the growth enhancement due to isotopic resonance was likely to be different for different bacteria phenotypes[34], the increase of the data spread with the size of the effect was expected. In other domains of growth measurements, the lag phase showed no significant change, while the maximum growth density had a maximum at 3%, but with a below-threshold significance (p = 0.085).

To validate the presence of the $^{15}$N resonance, a narrow range of $^{15}$N content, 2.0-4.1%, was investigated with a 0.3% step. The results (Figure 2b) confirmed the existence of a resonance around 3.5-3.8%. The size of the effect (a 0.3% increase in the maximum growth rate compared to 2.0% $^{15}$N content, p = 0.006) was consistent with that of the broad-range experiment. The other two growth parameters were also supportive of faster growth: the lag phase had a statistically significant minimum at 3.2%, and the maximum density was enhanced at 3.5% (Figures 2c, d). The combined p-value of these observations is <0.00005. Note that an association of the effect magnitude with the size of data spread increases the p-value; therefore, the above p-value is a conservative estimate.

*Testing the $^{13}$C ≈0.35% resonance*

The resonance for $^{13}$C predicted at 0.35% was tested at four different $^{13}$C concentrations in the range from 0.2% to 1.1% (normal terrestrial value). Figure 3 shows the results for the growth parameters. The presence of a resonance at 0.35% is supported by both maximum growth rate (+0.9%) and maximum density (+1.5%); the combined p-value is $10^{-6}$. At the same time, the lag phase decreased strongly with $^{13}$C content decrease, and reached a minimum (-3%) at 0.2% $^{13}$C.

To explore the effect of temperature on the growth rate enhancement under on-resonance condition, *E. coli* was grown in 0.35% $^{13}$C at a temperature ranging from 15 °C to 41 °C. The maximum growth rates were always higher than in isotopically normal media (the combined p-value is $10^{-25}$), with the largest increase, ≥1%, observed in the range between 25 °C and 35 °C (Figure 4).

One interpretation of the effect of isotopic resonance is that the complexity reduction leads to lower density of quantum-mechanical states, which is similar (but not equivalent) to a higher internal energy per degree of freedom, i.e. higher temperature[30]. In the range of 25-35

°C, the growth rate of *E. coli* is still sufficiently lower than the maximum achieved at ca. 39 °C, so that the analogue of temperature increase achieved via isotopic resonance could have a noticeable positive effect. At 39 °C, the isotopic resonance is no longer similar to temperature increase, but it still accelerates the growth, albeit less than at lower temperature.

*Testing the super-resonance ($^{13}C$ = 9.54%; $^{15}N$ = 10.89%; $^{18}O$ = 6.6%).*

As a first step, we tested the resonance at $^{18}O \approx 6.6\%$ that is valid for molecules containing H and O, i.e. water. The presence of a strong resonance (ca. -1.3%, p <0.0002) is obvious in the lag phase domain (Figure 5a). At the same time, both maximum growth rate and maximum density increase monotonously with $^{18}O$ content (Figures S2a and b), consistent with a strong effect on growth of bacteria that $^{18}O$ is known to induce[11].

Similarly, the $^{13}C$ resonance at 9.5% was tested, which should act on molecules with C and H, i.e. hydrocarbons. Unlike other elements, $^{13}C$ enrichment does not destroy the terrestrial resonance for Z = 0. Therefore, it was not expected that $^{13}C$ enrichment up to 10% would have a strong effect on bacterial growth. Indeed, both maximum growth rate and maximum density remained unchanged below 13% enrichment, while the lag phase was lower by 0.5% in the range of 8-13% $^{13}C$ (Figure S3).

At $^{15}N \approx 10.9\%$, there is a resonance acting on molecules with N and H, i.e. on ammonia, but no increase was observed for any growth parameter (Figure S4), with all values being statistically indistinguishable from controls. This was not surprising, given that the concentration of free ammonia in bacteria and growth media is low.

Pairwise enrichments combining the above resonances not surprisingly yielded strongest effect for C+O and smallest effect for C+N (Figure 5c). To compare the magnitudes of the effects for individual and combined isotopic enrichments, the effect sizes of the three growth parameters were added together (Figure 5c). Combined C+O and O+N enrichments gave larger effects than the combined effects of individual enrichments of C, O and N. But by far the largest effect was observed for the triple enrichment C+O+N (Figure 5b), which predicted to be a super-resonance for all CHON molecules. The maximum growth rate increased by 0.6%, while the maximum density by 3%, with the lag phase shortened by 2.4%. The overall effect of triple enrichment was larger than any combination of the individual and/or pairwise enrichments. Thus the super-resonance conditions provide an extremely comfortable environment for bacterial growth.

**Conclusions and Outlook**

Precise measurements of *E. coli* growth parameters at different isotopic compositions of $^{13}$C, $^{15}$N and $^{18}$O provided statistically significant confirmation for enhanced growth at a number of predicted resonance isotopic compositions. The temperature dependence of the resonance effect and the relative magnitude comply with expectations. An extremely strong in statistical terms effect was observed at a triple enrichment of $^{13}$C, $^{15}$N and $^{18}$O, which exceeded the combined effects of the individual or pairwise enrichments. Taken together, these observations leave no doubts in reality of the isotopic resonance phenomenon. A number of strong non-resonance effects were uncovered, such as that of $^{18}$O, where a measureable impact was detected starting from 3% enrichment.

These results, combined with our recent deuterium enrichment studies[33], and with a bulk of literature that has been published by others starting from 1930s[15–29], open a venue for scientific and industrial exploration of isotopic resonance in a whole range of fields. In astrobiology, the impact of the isotopic resonance phenomenon on the origin of life on Earth has to be seriously considered. The atmospheric isotopic compositions of other plants of our Solar system differ from that of Earth, especially in deuterium content[35], so that no strong resonance exists on Mars or Venus (Figure 6). If this factor is linked to the probability of life, as the hypothesis suggests, searching for life on exoplanets will have an additional narrowing to parameter to consider. In space exploration, growing food on the Moon and on other planets may be accelerated by "tuning up" the growth environment to a convenient isotopic resonance. In biotechnology, production of biomass and biomolecules may be boosted; in chemistry, organic and perhaps even inorganic synthesis may benefit as well. It remains to be tested whether isotopic resonance can increase the rate of highly exothermic reactions, such as combustion and explosion, but from first principles, it is likely. Food industry may also be affected, as stable isotopes are considered safe, especially at low enrichment[11]. Last but not least, medicine has already been exploring, although on a limited scale, the retardation of cancer cell growth at off-resonance conditions[28,29].

As a final word, stable isotopes remain one of the few relatively unexplored frontiers in life sciences and technology. The validation of the isotopic resonance phenomenon adds incentive to exploring this highly promising frontier in earnest.

**Acknowledgements**

This work was funded by the Swedish Research Council, grant 2011-3726 to RZ. The authors are grateful to Tatyana Perlova and Alexander R. Zubarev for creation of visualization tools for isotopic resonance calculations.

**References**

1.  Croswell, K. *The Alchemy of the Heavens: Searching for Meaning in the Milky Way.* (Knopf Doubleday Publishing Group, 1996).

2.  Cameron, A. G. W. Abundances of the elements in the solar system. *Space Sci. Rev.* **15,** 121–146 (1970).

3.  Morgan, J. W. & Anders, E. Chemical composition of Earth, Venus, and Mercury. *Proc. Natl. Acad. Sci.* **77,** 6973–6977 (1980).

4.  Frieden, E. The chemical elements of life. *Sci. Am.* **227,** 52–60 (1972).

5. Urey, H. C., Brickwedde, F. G. & Murphy, G. M. A hydrogen isotope of mass 2. *Phys. Rev.* **39,** 164–166 (1932).

6. Katz, J. J., Crespy, H. L., Hasterlik, R. J., Thomson, J. F. & Finkel, A. J. Some observations on biological effects of deuterium, with special reference to effects on neoplastic processes. *J. Natl. Cancer Inst.* **18,** 641–659 (1957).

7. Czajka, D. M., Finkel, A. J., Fischer, C. S. & Katz, J. J. Physiological effects of deuterium on dogs. *Am. J. Physiol.* **201,** 357–362 (1961).

8. Katz, J. J. & Crespi, H. L. Deuterated organisms: Cultivation and uses. *Science* **151,** 1187–1194 (1966).

9. Uphaus, R. A., Flaumenhaft, E. & Katz, J. J. A living organism of unusual isotopic composition. Sequential and cumulative replacement of stable isotopes in Chlorella vulgaris. *Biochim. Biophys. Acta* **141,** 625–632 (1967).

10. Gregg, C. T., Hutson, J. Y., Prine, J. R., Ott, D. G. & Furchuer, J. E. Substantial replacement of mammalian body carbon with carbon-13. *Life Sci.* **13,** 775–782 (1973).

11. Klein, P. D. & Klein, E. R. Stable isotopes: origins and safety. *J. Clin. Pharmacol.* **26,** 378–382 (1986).

12. Filiou, M. D. *et al.* Proteomics and metabolomics analysis of a trait anxiety mouse model reveals divergent mitochondrial pathways. *Biol. Psychiatry* **70,** 1074–1082 (2011).

13. Filiou, M. D. *et al.* The 15N isotope effect in Escherichia coli: A neutron can make the difference. *Proteomics* **12,** 3121–3128 (2012).

14. Shchepinov, M. S. Do "heavy" eaters live longer? *BioEssays* **29,** 1247–1256 (2007).

15. Barnes, T. C. A possible physiological effect of the heavy isotope of H in water. *J. Am. Chem. Soc.* **55,** 4332–4333 (1933).

16. Barnes, T. C. & Larson, E. J. Further experiments on the physiological effect of heavy water and of ice water. *J. Am. Chem. Soc.* **55,** 5059–5060 (1933).

17. Barnes, T. & Larson, E. The influence of heavy water of low concentration on Spirogyra, Planaria and on enzyme action. *Protoplasma* **22,** 431–443 (1935).

18. Barnes, T. C. The effect of heavy water of low concentration on Euglena. *Science* **79,** 370 (1934).

19. Richards, O. W. The effect of deuterium on the growth of yeast. *J. Bacteriol.* **28,** 289–294 (1934).

20. Richards, O. W. The growth of yeast in water containing deuterium. *Am. Jour. Bot.* **20,** 679–680 (1933).

21. Lockemann, G. & Leunig, H. Über den Einfluß des "schweren Wassers" auf die biologischen Vorgänge bei Bakterien. *Berichte der Dtsch. Chem. Gesellschaft* **67,** 1299–1302 (1934).

22. Macht, D. I. & Davis, M. E. Some pharmacological experiments with deuterium. *J. Am. Chem. Soc.* **56,** 246 (1934).

23. Curry, J., Pratt, R. & Trelease, S. F. Does dilute heavy water influence biological processes? *Science* **81,** 275–277 (1935).

24. Lobyshev, V. I., Tverdislov, V. A., Vogel, J. & Iakovenko, L. V. Activation of Na,K-ATPase by small concentrations of D2O, inhibition by high concentrations. *Biofizika* **23,** 390–391 (1978).

25. Lobyshev, V. I., Fogel' Iu, Iakovenko, L. V., Rezaeva, M. N. & Tverdislov, V. A. D2O as a modifier of ionic specificity of Na, K-ATPase. *Biofizika* **27,** 595–603 (1982).

26. Lobyshev, V. I. Activating influence of heavy water of small concentration on the regeneration of hydroid polyp obelia geniculata. *Biofizika* **28,** 666–668 (1983).

27. Somlyai, G. *et al.* Naturally occurring deuterium is essential for the normal growth rate of cells. *FEBS Lett.* **317,** 1–4 (1993).

28. Krempels, K., Somlyai, I. & Somlyai, G. A retrospective evaluation of the effects of deuterium depleted water consumption on 4 patients with brain metastases from lung cancer. *Integr. Cancer Ther.* **7,** 172–181 (2008).

29. Krempels, K. *et al.* A retrospective study of survival in breast cancer patients undergoing deuterium depletion in addition to conventional therapies. *J. Cancer Res. Ther.* **1,** 194–200 (2013).

30. Zubarev, R. A. *et al.* Early life relict feature in peptide mass distribution. *Cent. Eur. J. Biol.* **5,** 190–196 (2010).

31. Artemenko, K. a *et al.* Two dimensional mass mapping as a general method of data representation in comprehensive analysis of complex molecular mixtures. *Anal. Chem.* **81,** 3738–3745 (2009).

32. Zubarev, R. A. Role of stable isotopes in life-testing isotopic resonance hypothesis. *Genomics Proteomics Bioinformatics* **9,** 15–20 (2011).

33. Xie, X. & Zubarev, R. A. Effects of low-level deuterium enrichment on bacterial growth. *PLoS One* **9,** e102071 (2014).

34. Nikitin, D. I., Oranskaya, M. N. & Lobyshev, V. I. Specificity of bacterial response to variation of isotopic composition of water. *Biofizika* **48,** 678–682 (2003).

35. Saal, A. E., Hauri, E. H., Van Orman, J. A. & Rutherford, M. J. Hydrogen isotopes in lunar volcanic glasses and melt inclusions reveal a carbonaceous chondrite heritage. *Science* **340,** 1317–1320 (2013).

36. Fegley, B. in *Glob. Earth Phys*. (ed. Ahrens, T. J.) 320–345 (American Geophysical Union, Washington, D. C., 2013).

**FIGURES**

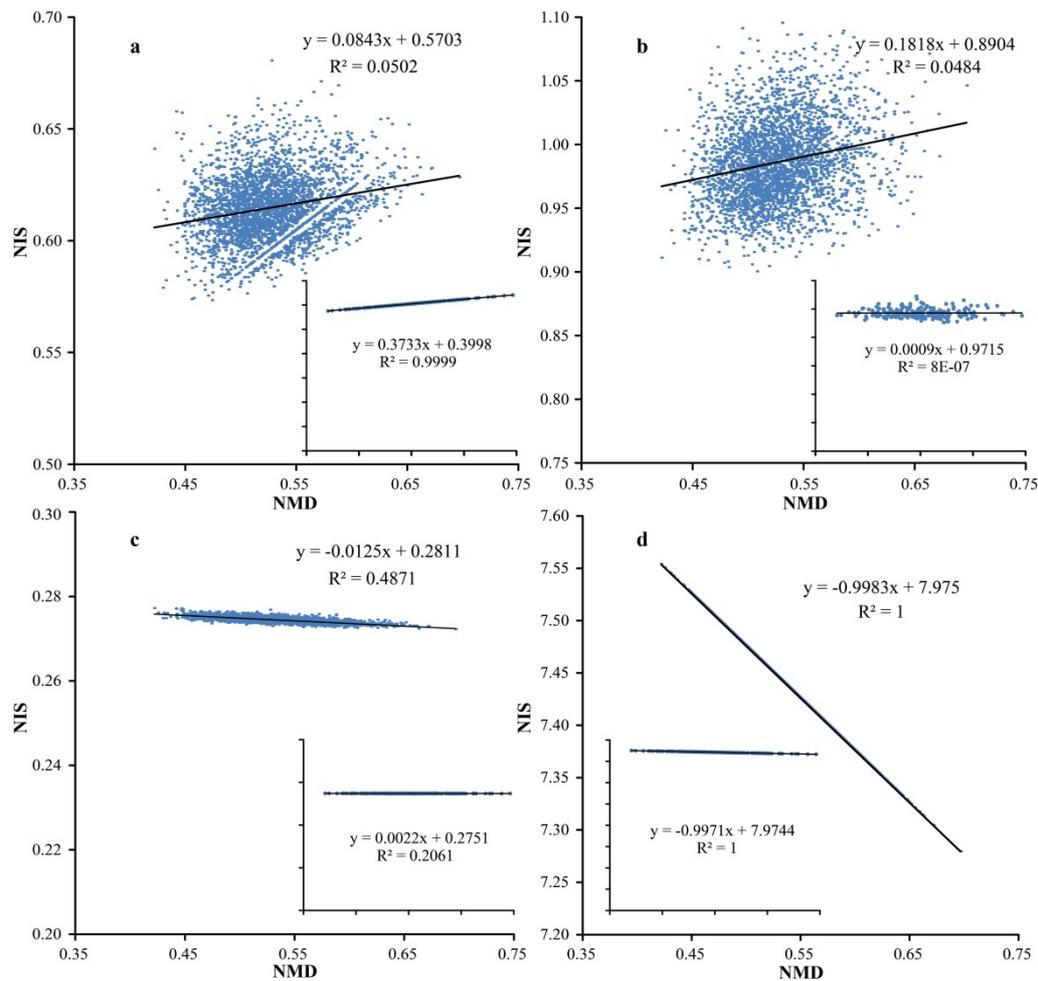

**Figure 1.** 2D mass plots of 3,000 *E. coli* tryptic peptides at different isotopic ratios of CHON. Insets show only peptides with Z = 0. The axes represent: (x) normalized monoisotopic defect (NMD), and (y) normalized isotopic shift (NIS). (a) Terrestrial isotopic ratios; the gap with a central line correspond to the terrestrial isotopic resonance for Z = 0 molecules. (b) Zero-slope resonance at ≈3.5% $^{15}N$ for Z = 0 molecules. (c) Zero-slope resonance at $^{13}C$ ≈0.35% for Z = 0 molecules and a near-resonance for all molecules. (d) The "super-resonance" at $^{13}C$ ≈9.5%, $^{15}N$ ≈10.9% and $^{18}O$ ≈6.6% for all molecules.

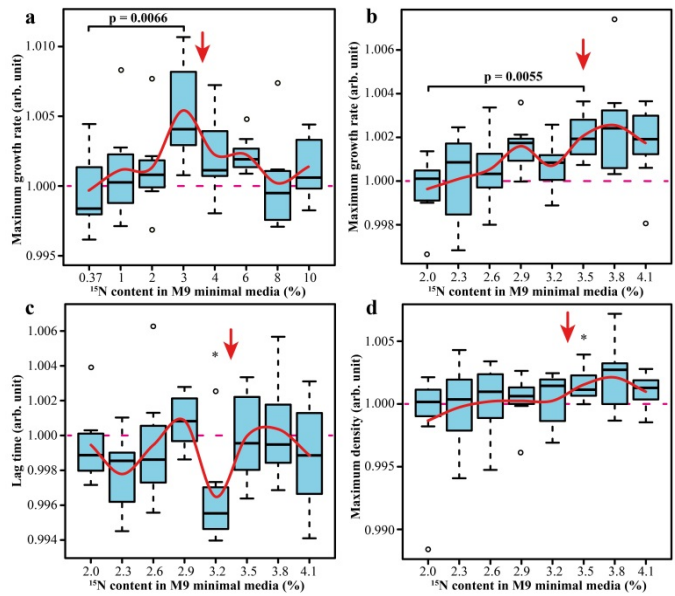

**Figure 2:** Growth parameters of *E. coli* grown in M9 minimal media with varying composition of $^{15}$N: (a), (b) – maximum growth rate; (c) – lag time; (d) – maximum density. * denotes p<0.05.

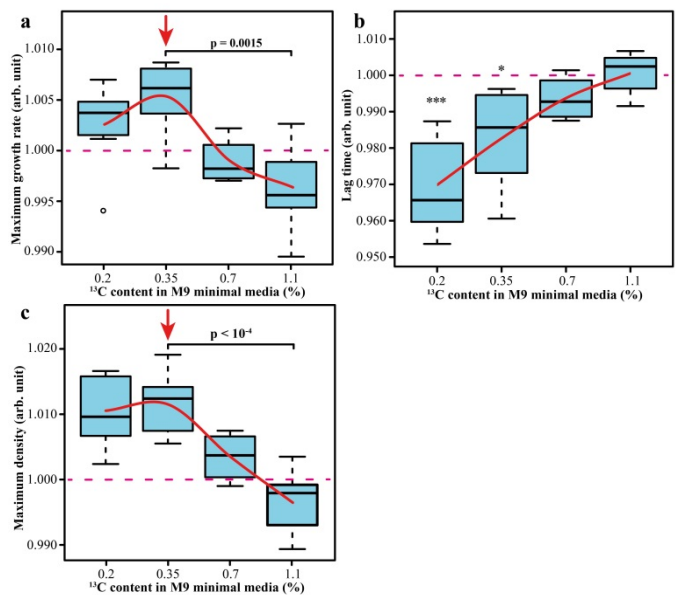

**Figure 3:** Growth parameters of *E. coli* grown in M9 minimal media with varying composition of $^{13}$C: (a) – maximum growth rate; (b) – lag time; (c) – maximum density. * denotes p<0.05.

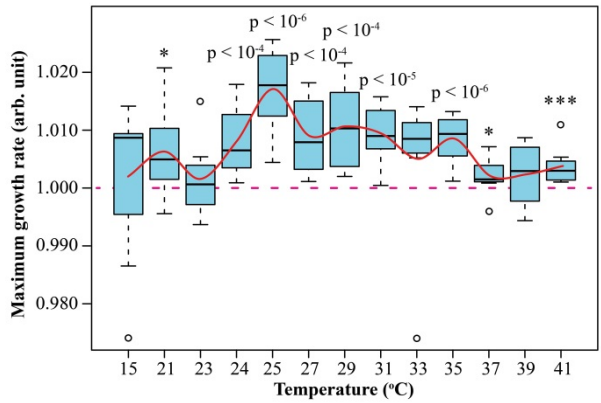

**Figure 4.** Maximum growth rate of *E. coli* grown in M9 minimal media with 0.35% of $^{13}$C at different temperatures. * denotes p<0.05, ** denotes p<0.05, etc.

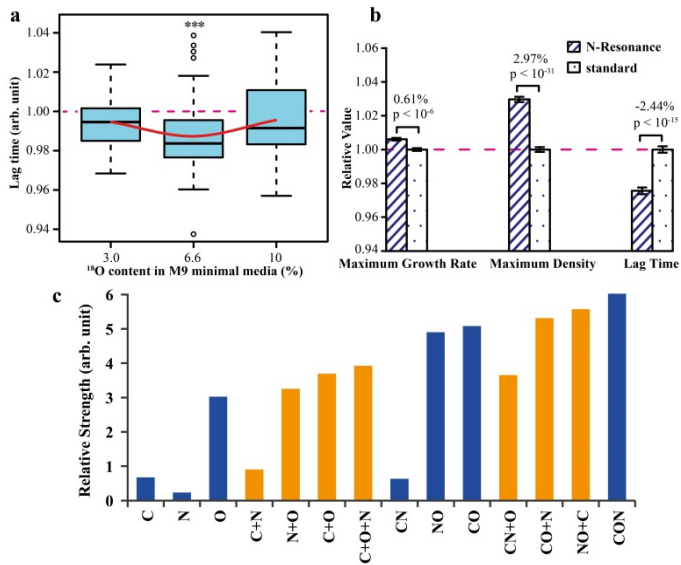

**Figure 5.** (a) Resonance in lag time at 6.6% $^{18}$O. (b) Resonance at the "super-resonance": at $^{13}$C ≈9.5%, $^{15}$N ≈10.9% and $^{18}$O ≈6.6%. (c) Relative magnitudes of the effects of individual and combined isotope enrichment: blue columns – experimental results; orange columns – extrapolated data. * denotes p<0.05, ** denotes p<0.05, etc.

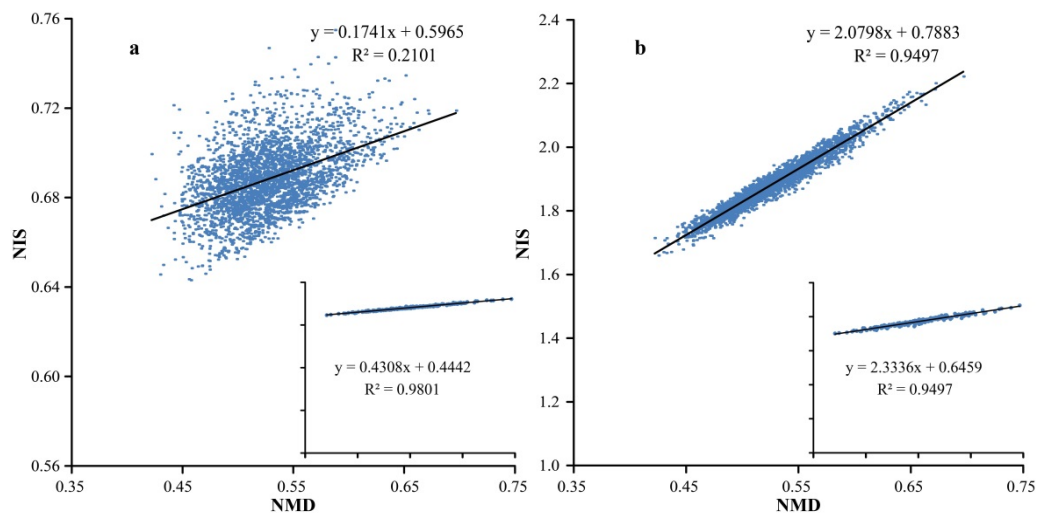

**Figure 6.** Same as in Figure 1 but for atmospheric isotopic compositions[36] of (a) Mars, (b) Venus.

**Methods**

*Resonance prediction.* The position and, to a certain degree, relative strength of the resonance is predicted to relate to the abundance of molecules and their role it affects, as well as to the degree of complexity reduction it induces. The terrestrial resonance in Figure 1a concerns molecules (not necessarily polypeptides or amino acids) following the rule: $Z = 0$, where $Z = C - (N+H)/2$ [30]. The linear correlation between NMD and NIS at this resonance has a non-zero slope, which means the lowest degree of complexity reduction: out of the 14 parameters needed to describe the average molecular mass of a CHON molecule, six factors remain at resonance conditions: four monoisotopic masses and two parameters of the line. Only a minor fraction of all molecules satisfies the rule $Z = 0$, and thus the terrestrial resonance should be relatively weak. Thus this resonance could be easily improved or superseded, either by "tuning up" the CHON isotopic abundances, e.g. to 0.03-0.05% D, or by changing the line slope to zero. The zero slope can be achieved by $^{15}$N enrichment to ≈3.5% (Figure 1b), even though the line becomes more diffuse. In contrast, the resonance at $^{13}$C $\approx 0.35\%$ not only gives a zero slope for $Z = 0$, but a near-zero slope for all other molecules. The corresponding line in Figure 1c, while not perfect, is much less spread from the line compared to the "galaxy" in Figure 1a. Thus this resonance should be stronger than the $^{15}$N $\approx 3.5\%$ resonance. The resonance at $^{18}$O $\approx 6.6\%$ is of different nature: the average isotopic masses become proportional to the nominal (integer) masses, which totally excludes the monoisotopic masses from the equation. Such complexity reduction is difficult to supersede. The resonance at $^{18}$O $\approx 6.6\%$ affects only species composed of hydrogen and oxygen, but that includes water, the most ubiquitous and important for life molecule. Additional enrichment of other isotopes to fulfil the same resonance condition, i.e. either $^{13}$C to 10.9% or $^{15}$N to 9.5% or both isotopes together (Figure 1d), should increase the effect further.

***Chemicals and materials.*** The bacteria were grown in M9 minimal media, with the isotopic composition varied by mixing normal ingredients with $^{12}$C- or $^{13}$C-glucose, D$_2$O, H$_2$$^{18}$O or $^{15}$NH$_4$Cl, keeping the molecular composition of the media constant. *See supplementary materials.*

***Sample preparation*** – *See supplementary materials.*

***E. coli growth measurements*** – *See supplementary materials.*

***Data analysis*** – *See supplementary materials.*

**Supplementary Materials**

**Chemicals and materials**

Glycerol stock of *E. coli* BL 21 strain (stored at -80 $^o$C) was obtained from the micorbiology lab of the department of Medical Biochemistry and Biophysics, Karolinska Institutet, Stockholm. Chemicals used to prepare M9 minimal media, including D-glucose ($C_6H_{12}O_6$, normal isotopic composition, 1.1% $^{13}$C, 180.16 g/mol), disodium hydrogen phosphate ($Na_2HPO_4 \cdot 2H_2O$), monopotassium phosphate ($KH_2PO_4$), sodium chloride (NaCl), magnesium sulfate ($MgSO_4$), calcium chloride ($CaCl_2$), ammonia chloride ($NH_4Cl$, normal isotopic composition, 0.37% $^{15}$N, 53.49 g/mol), were purchased from Sigma-Aldrich (Schnelldorf, Germany). Ammonia chloride containing 99.16% $^{15}$N ($^{15}NH_4Cl$, 54.49 g/mol) and D-glucose containing 99% $^{13}$C ($^{13}C_6H_{12}O_6$, 186.11 g/mol) were purchased from Silantes (München, Germany). D-glucose depleted with $^{13}$C ($^{12}C_6H_{12}O_6$, 99.9% $^{12}$C, 180.09 g/mol) was purchased from Cambridge Isotope Laboratories (Andover, MA, USA). Water ($H_2^{18}O$) enriched with 97% $^{18}$O was purchase from Sigma-Aldrich (Schnelldorf, Germany). Select agar was purchased from Invigen (Paisley, UK). Distilled water was prepared with a Milli-Q device from Millipore (Billerica, MA, USA).

Vacuum filtration system with 0.2 μm polyethersulfone (PES) membrane for bacteria media sterilization was purchased from VWR (Stockholm, Sweden). Petri dishes (90 x 15 mm), inoculating loops and corning sterile culture tubes (16 x 125 mm) were purchased Sigma-Aldrich. Sterile plastic conical tubes (50 mL and 15 mL) for sample preparation were purchased from Sarstedt (Nümbrecht, Germany). The BioScreen C automatic fermentor was obtained from Oy Growth Curves AB Ltd (Helsinki, Finland).

**Sample preparation**

*(1) To grow E. coli at different content of $^{15}N$.*

*M9 minimal media preparation*

<u>5-time concentrated M9 minimal salts</u> stock solution was prepared by dissolving 42.5 g Na$_2$HPO$_4 \cdot$ 2H$_2$O, 15 g KH$_2$PO$_4$ and 2.5 g NaCl in Milli-Q water to a final volume of 1000 mL. The solution was then sterilized by autoclaving and stored at 4 $^{\circ}$C for further use. To prepare M9 minimal media, the salts stock solution was diluted five times with Milli-Q water.

<u>Nitrogen-free M9 minimal media</u> (without nitrogen source $^{15}$NH$_4$Cl) were prepared by mixing the following components: 800 mL Milli-Q water, 200 mL M9 concentrated salts stock solution, 2 mL of 1 M MgSO$_4$ solution, 0.1 mL of 1 M CaCl$_2$ solution, 5 g D-glucose.

<u>M9 minimal media with normal isotopic composition (0.37% $^{15}$N, 99.63% $^{14}$N)</u> were prepared by dissolving 200.00 mg of NH$_4$Cl (normal isotopic composition) in 200.0 g of nitrogen-free M9 minimal media solution weighed in a sterile plastic bottle. The prepared media were then sterilized by filtering with a 200 mL vacuum filtration system through a 0.2 μm PES membrane.

<u>M9 minimal media enriched with $^{15}$N (99.16% $^{15}$N, 0.84% $^{14}$N)</u> were prepared by dissolving 203.74 mg of $^{15}$NH$_4$Cl (99.16% $^{15}$N, 0.84% $^{14}$N) in 200.0 g nitrogen-free M9 minimal media solution, followed by filtering with a 250 mL vacuum filtration system through a 0.2 μm PES membrane.

*Preparation of streak agar plates*

M9 minimal media agar plates with normal isotopic composition (0.37% $^{15}$N, 99.63% $^{14}$N) were prepared by dissolving 3 g of agar powder in 200 mL M9 minimal media (0.37% $^{15}$N). The obtained mixture was sterilized by autoclaving, then cooled down to ca. 60 °C and finally poured into Petri dishes (ca. 15 mL agar solution per plate). The agar plates were allowed to solidify at room temperature for ca. 10 mins, sealed with parafilm and stored at 4 °C till further use.

*E. coli* streak agar plates were prepared by streaking *E. coli* from -80 °C glycerol stock onto M9 minimal media agar plate followed by ca. 40-hour incubation at 37 °C to form visible isolated colonies. Streak agar plates were stored at 4 °C for the following experiments no longer than one week. Fresh steak agar plates were prepared regularly once a week.

*Preparation of E. coli sample on honeycomb well plate*

From the *E. coli* agar plate, one isolated colony was picked with a sterile loop into 5 mL M9 minimal media (normal isotopic condition) and incubated at 37 °C while shaking 250 r.p.m for 5-6 hours until it reached its early exponential phase with optical density (O.D.) around 0.2, measured with Colorimeter WPA CO75 (York, UK).

Sample preparation workflow is shown in Figure S5. In each experiment, three stock solutions were used. Stock A for preparing sample $S_A$, and stock B for preparing sample $S_B$ were obtained by mixing M9 minimal media with normal isotopic condition (0.37% of $^{15}$N) and $^{15}$N enriched M9 minimal media (99.16% of $^{15}$N) at certain ratio (Table 1). M9 minimal media with normal isotopic condition (0.37% of $^{15}$N) were used to prepare stock solution of standard A and standard B. The final solutions were dispensed into the honeycomb well plates using programmed robotic system (Tecan, Genesis RSP 150, Männedorf, Switzerland).

A 20 μL (36 μL for 50% of $^{15}$N) aliquot of the incubated *E. coli* culture (O.D. ≈0.2) was diluted in 45 mL M9 minimal media of normal isotopic composition (0.37% $^{15}$N) prepare diluted *E. coli* culture for robot sample preparation. First, 300 μL M9 minimal media without bacteria were introduced into each of the border wells on both plates A and B to serve as blanks (no color code in Figure S6[33]). Second, 30 μL (151 μL for 50% of $^{15}$N) aliquot of stock A was dispensed into each "sample" well on plate $P_A$ (marked with purple in Figure S6) to prepare 32 replicates of sample $S_A$. Third, 30 μL (151 μL for 50% of $^{15}$N) aliquot of stock B was dispensed to each "sample" well on plate $P_B$ (marked with blue in Figure S6), resulting in 32 replicates of sample B. Fourth, 30 μL (151 μL for 50% of $^{15}$N) aliquot of stock solution of standards (0.37% $^{15}$N) was added into each "standard" well on plate $P_A$ and $P_B$ (marked with yellow in Figure S6) to prepare 32 reference standards on each plate. Finally, 270 μL (149 μL for 50% of $^{15}$N) of the diluted *E. coli* culture was dispensed into each well except blank wells. In total, 32 replicates pairs of "sample" and "standard" wells were prepared on each plate.

**Table 1. Stock solutions and their corresponding $^{15}$N compositions in the final samples.** Stock solution (column one) was prepared by mixing M9 minimal media with normal isotopic composition (0.37% $^{15}$N, column two) and M9 minimal media enriched with 99.16% $^{15}$N (column three) at certain ratio, resulting in the final $^{15}$N composition in the sample (column four).

| $^{15}$N composition in stock solution | M9 minimal media enriched with 0.37% $^{15}$N (μL) | M9 minimal media enriched with 99.16% $^{15}$N (μL) | $^{15}$N composition in the final sample (honeycomb well plate) |
|---|---|---|---|
| 0.37% | 5000 | 0 | 0.37% |
| 6.67% | 4991 | 340 | 1% |
| 16.67% | 4185 | 827 | 2% |
| 26.67% | 3856 | 1399 | 3% |
| 36.67% | 3195 | 1856 | 4% |
| 56.67% | 2200 | 2915 | 6% |
| 76.67% | 1179 | 4000 | 8% |
| 96.67% | 129 | 4991 | 10% |
| 16.67% | 4175 | 825 | 2.0% |
| 19.67% | 4090 | 993 | 2.3% |
| 22.67% | 3883 | 1132 | 2.6% |
| 25.67% | 3776 | 1300 | 2.9% |
| 28.67% | 3614 | 1451 | 3.2% |
| 31.67% | 3491 | 1619 | 3.5% |
| 34.67% | 3311 | 1761 | 3.8% |
| 37.67% | 3152 | 1912 | 4.1% |
| 99.16% | 0 | 16,000 | 50% |

*(2) To grow E. coli at different content of $^{13}C$.*

*M9 minimal media preparation*

Carbon-free M9 minimal media (without D-glucose) were prepared by mixing the following components: 800 mL Milli-Q water, 200 mL M9 concentrated salts stock solution, 2 mL of 1 M $MgSO_4$ solution, 0.1 mL of 1 M $CaCl_2$ solution, 1 g $NH_4Cl$.

M9 minimal media with normal isotopic composition (1.1% $^{13}C$, 98.9% $^{12}C$) were prepared by dissolving 2000.78 mg of D-glucose (normal isotopic composition) in 400.0 g of carbon-free M9 minimal media solution weighed in a sterile plastic bottle. The prepared media were then sterilized by filtering with a 500 mL vacuum filtration system through a 0.2 μm PES membrane.

M9 minimal media depleted with $^{13}$C (0.1% $^{13}$C, 99.9% $^{12}$C) were prepared by dissolving 500.00 mg of D-glucose (0.1% $^{13}$C, 99.9% $^{12}$C) in 100.0 g carbon-free M9 minimal media solution, followed by filtering with a 250 mL vacuum filtration system through a 0.2 μm PES membrane.

M9 minimal media enriched with $^{13}$C (99% $^{13}$C, 1% $^{12}$C) were prepared by dissolving 2066.86 mg of D-glucose (99% $^{13}$C, 1% $^{12}$C) in 400.0 g carbon-free M9 minimal media solution, followed by filtering with a 500 mL vacuum filtration system through a 0.2 μm PES membrane.

*Preparation of E. coli sample on honeycomb well plate*

Sample preparation workflow is shown in Figure S5. In each experiment, three stock solutions were used. Stock A for preparing sample $S_A$, and stock B for preparing sample $S_B$ were obtained by mixing M9 minimal media at normal isotopic condition (1.1% of $^{13}$C) with $^{13}$C depleted minimal media (0.1% $^{13}$C) or $^{13}$C enriched M9 minimal media (99% of $^{13}$C) at certain ratio (Table 2 and Table 3). M9 minimal media at normal isotopic condition (1.1% of $^{13}$C) were used to prepare stock solution of standard A and standard B. The final solutions were dispensed into the honeycomb well plates using the Tecan robot or pipettes.

To test *E. coli* growth at 0.1-1.1% $^{13}$C, stock solutions were prepared according to Table 2. A 40 μL aliquot of the incubated *E. coli* culture (O.D. ≈0.2) was diluted in 10 mL M9 minimal media (normal) to prepare diluted *E. coli* culture. To minimize the cost of $^{13}$C depleted media, pipettes were used for part of the sample preparation here. First, 400 μL M9 minimal media without bacteria were introduced into each of the border wells on both plates A and B to serve as blanks (no color code in Figure S6). Second, 360 μL aliquot of stock A was dispensed manually by pipette into each "sample" well on plate $P_A$ (marked with purple in Figure S6) to prepare 32 replicates of sample $S_A$. Third, 360 μL aliquot of stock B was dispensed to each "sample" well on plate $P_B$ (marked with blue in Figure S6), resulting in 32 replicates of sample B. For the next step, 360 µL aliquot of stock solution of standards (1.1% $^{13}$C) was added into each "standard" well on plate $P_A$ and $P_B$ (marked with yellow in Figure S6) to prepare 32 reference standards on each plate. Finally, 40 µL of the diluted *E. coli* culture was dispensed into each well except blank wells with robot. In total, 32 replicates pairs of "sample" and "standard" wells were prepared on each plate.

To test *E. coli* growth at 3-13% $^{13}$C, stock solutions were prepared according to Table 3. A 20 µL aliquot of the incubated *E. coli* culture (O.D. ≈0.2) was diluted in 45 mL M9 minimal media (normal) to prepare diluted *E. coli* culture for robot sample preparation. First, 300 µL M9 minimal media without bacteria were introduced into each of the border wells on both plates A and B to serve as blanks (no color code in Figure S6). Second, 40 µL aliquot of stock A was dispensed into each "sample" well on plate $P_A$ (marked with purple in Figure S6) to prepare 32 replicates of sample $S_A$. Third, 40 µL aliquot of stock B was dispensed to each "sample" well on plate $P_B$ (marked with blue in Figure S6), resulting in 32 replicates of sample B. For the next step, 40 µL aliquot of stock solution of standards (1.1% $^{13}$C) was added into each "standard" well on plate $P_A$ and $P_B$ (marked with yellow in Figure S6) to prepare 32 reference standards on each plate. Finally, 260 µL of the diluted *E. coli* culture was dispensed into each well except blank wells. In total, 32 replicates pairs of "sample" and "standard" wells were prepared on each plate.

**Table 2. Stock solutions and their corresponding $^{13}$C compositions in the final samples.** Stock solution (column one) was prepared by mixing M9 minimal media depleted with $^{13}$C (0.1% $^{13}$C, column two) and M9 minimal media with normal isotopic composition (1.1% $^{13}$C, column three) at certain ratio, resulting in the final $^{13}$C composition in the sample (column four).

| $^{13}$C composition in stock solution | M9 minimal media with 0.1% $^{13}$C (μL) | M9 minimal media with 1.1% $^{13}$C (μL) | $^{13}$C composition in the final sample (honeycomb well plate) |
|---|---|---|---|
| 0.1% | 15040 | 0 | 0.2% |
| 0.2667% | 12450 | 2490 | 0.35% |
| 0.6556% | 6640 | 8300 | 0.7% |
| 1.1% | 0 | 15040 | 1.1% |

**Table 3. Stock solutions and their corresponding $^{13}$C compositions in the final samples.**

Stock solution (column one) was prepared by mixing M9 minimal media with normal isotopic composition (1.1% $^{13}$C, column two) and M9 minimal media enriched with $^{13}$C (99% $^{13}$C, column three) at certain ratio, resulting in the final $^{13}$C composition in the sample (column four).

| $^{13}$C composition in stock solution | M9 minimal media with 1.1% $^{13}$C (µL) | M9 minimal media with 99% $^{13}$C (µL) | $^{13}$C composition in the final sample (honeycomb well plate) |
|---|---|---|---|
| 15.35% | 5923 | 1009 | 3% |
| 37.85% | 4288 | 2577 | 6% |
| 52.85% | 3280 | 3678 | 8% |
| 64.40% | 2422 | 4431 | 9.54% |
| 75.35% | 1634 | 5130 | 11% |
| 90.35% | 604 | 6232 | 13% |

**(3) *To grow E. coli at different content of $^{18}$O***

*Preparation of E. coli sample on honeycomb well plate*

In each experiment, four stock solutions were used. Stock A for preparing sample S$_A$, and stock B for preparing sample S$_B$ were obtained by mixing M9 minimal media at normal isotopic condition with sterile $^{18}$O water (97% $^{18}$O) at certain ratio (Table 4). For the preparation of stock solutions of standard A and standard B, M9 minimal media were mixed with sterile Milli-Q water at the same ratio as stock A and stock B. The final solutions were dispensed into the honeycomb well plates using the Tecan robot.

A 21 μL aliquot of the incubated *E. coli* culture (O.D. ≈0.2) was diluted in 35 mL M9 minimal media to prepare diluted *E. coli* culture for following sample preparation. First, 300 μL M9 minimal media without bacteria were introduced with robot into each of the border wells ("edge cells") on both plates $P_A$ and $P_B$ (72 wells in total) to serve as blank samples (no color code in Figure S6) with robot. Second, 100 μL aliquot of stock A was dispensed into each "sample" well (marked with purple on plate $P_A$ in Figure S6) to prepare 32 replicates of sample $S_A$ manually by pipette. Third, 100 aliquot of stock solution of standard A was added manually by pipette into each "standard" well (marked with yellow on plate $P_A$ in Figure S6) to prepare 32 reference standards on plate. In the same way, wells were filled on plate $P_B$. Finally, 200 μL of the diluted *E. coli* culture was dispensed into each well except blank wells with robot. In total, 32 replicate pairs of "sample" and "standard" wells were prepared on each plate.

**Table 4. Stock solutions and their corresponding $^{18}$O compositions in the final samples.** Stock solution (column one) was prepared by mixing M9 minimal media with normal isotopic composition (0.2% $^{18}$O, column two) and M9 minimal media enriched with $^{18}$O (97% $^{18}$O, column three) at certain ratio, resulting in the final $^{18}$O composition in the sample (column four).

| $^{18}$O composition in stock solution | M9 minimal media (0.2% $^{18}$O, μL) | $^{18}$O water (99% $^{18}$O, μL) | $^{18}$O content in the final sample (honeycomb well plate) |
|---|---|---|---|
| 8.6% | 3536 | 336 | 3% |
| 19.4% | 3104 | 768 | 6.6% |
| 29.6% | 2696 | 1176 | 10% |

*(4) To grow E. coli at $^{13}$C ≈9.5%, normal isotopic composition of D, $^{18}$O ≈6.6% and $^{15}$N ≈10.9%.*

*M9 minimal media preparation*

Carbon and nitrogen free M9 minimal media (without carbon source and nitrogen source) were prepared by mixing the following components: 800 mL Milli-Q water, 200 mL M9 concentrated salts stock solution, 2 mL of 1 M MgSO$_4$ solution, and 0.1 mL of 1 M CaCl$_2$ solution.

Carbon-free M9 minimal media with 0.37% of $^{15}$N were prepared by dissolving 545.00 mg NH$_4$Cl (0.37% $^{15}$N) into 500.0 g carbon and nitrogen free M9 minimal media.

Carbon-free M9 minimal media with 99.16% of $^{15}$N were prepared by dissolving 555.19 mg NH$_4$Cl (99.16% of $^{15}$N) into 500.0 g carbon and nitrogen free M9 minimal media.

M9 minimal media with 1.1% of $^{13}$C and 0.37% of $^{15}$N were prepared by dissolving 1362.50 mg D-glucose (normal isotopic composition) into 250.0 g carbon-free M9 minimal media with 0.37% $^{15}$N. The prepared media were then sterilized by filtering with a 250 mL vacuum filtration system through a 0.2 µm PES membrane.

M9 minimal media with 26.4% of $^{13}$C and 31.9% of $^{15}$N were prepared by dissolving 1010.12 mg D-glucose (1.1% of $^{13}$C) and 364.02 mg of D-glucose (99% of $^{13}$C) into 250 g carbon-free M9 minimal media with 31.93% of $^{15}$N which were prepared by mixing 170.1 g carbon-free M9 minimal media with 0.37% $^{15}$N and 79.9 g carbon-free M9 minimal media with 99.16% $^{15}$N, followed by filtering with a 250 mL vacuum filtration system through a 0.2 µm PES membrane.

*Preparation of E. coli sample on honeycomb well plate*

Stock A: M9 minimal media with 26.4% of $^{13}$C, 31.9% of $^{15}$N and 19.4% of $^{18}$O were prepared by mixing 860 µL sterile $^{18}$O water (97% of $^{18}$O) with 3475.8 µL M9 minimal media with 26.4% of $^{13}$C and 31.9% of $^{15}$N (prepared above).

Stock solution for Standards: M9 minimal media with normal isotopic composition (1.1% of $^{13}$C, 0.37% of $^{15}$N, 0.2% of $^{18}$O) were prepared by mixing 860 µL sterile Milli-Q water with 3475.8 µL M9 minimal media with 1.1% of $^{13}$C and 0.37% of $^{15}$N (prepared above).

A 10 µL aliquot of the incubated *E. coli* culture (O.D. ≈0.2) was diluted into 15 mL M9 minimal media (1.1% of $^{13}$C and 0.37% of $^{15}$N) to prepare diluted *E. coli* culture for following sample preparation. First, 300 µL M9 minimal media without bacteria were introduced into each of the border wells ("edge cells") on $P_A$ serve as blank samples (no color code in Figure S6) with Robot. Second, 100 μL aliquot of stock A was dispensed into each "sample" well (marked with purple on plate $P_A$ in Figure S6) to prepare 32 replicates of sample $S_A$ manually by pipette. Third, 100 aliquot of stock solution of standards was added manually by pipette into each "standard" well (marked with yellow on plate $P_A$ in Figure S6) to prepare 32 reference standards on plate. Finally, 200 μL of the diluted *E. coli* culture was dispensed into each well except blank wells with the robot. In total, 32 replicate pairs of "sample" and "standard" wells were prepared on the plate.

## *E. coli* growth measurements

After sample preparation on the honeycomb well plate, *E. coli* concentration in each well was continuously monitored by measuring turbidity (with wide band filter 420–580 nm) using BioScreen C instrument with continuous shaking at 39 °C. Turbidity was sampled every six minutes and was monitored for ca. 22 hours to obtain a raw growth curve.

## Data analysis

Data analysis[33] was performed with Excel software as described in reference 33.

Using Microsoft Excel, the logarithm of turbidity was plotted against time. The slope for every 8-h interval was calculated, and the maximum value was determined. The extrapolation of the line with maximum slope to the background level of turbidity gave the lag time. The maximum turbidity for each replicate was taken as the maximum density. The obtained three values for each growth curve were treated in the same way as below.

For each "sample" A, the obtained value was normalized by that of the "standard" B. To minimize the influence of nonstatistical outliers that could arise due to gross errors in sample preparation and handling (e.g. differences in the geometry of the honeycomb wells, position-dependent sensitivity of the BioScreen C detector, etc.), the 32 replicates were divided into 4 groups according to their positions on the honeycomb well plate (group 1: columns 1 and 2; …, group 4: columns 7 and 8). In each group, the median of the eight values was calculated and then the four medians were averaged to obtain the value for a given plate and its standard deviation.

Altogether, seven independent 32-replicate experiments were performed for each $^{15}$N content point for the experiment to test *E. coli* growth at 0.37-10% $^{15}$N; seven independent 32-replicate experiments were performed for each $^{15}$N content point for the experiment to test *E. coli* growth at 2.0-4.1% $^{15}$N; seven independent 32-replicate experiments were performed for each $^{13}$C content point for the experiment to test *E. coli* growth at 0.1-1.1% $^{13}$C; five independent 32-replicate experiments were performed for each $^{13}$C content point for the experiment to test *E. coli* growth at 3-13% $^{13}$C and three independent experiments were performed to test *E. coli* growth at the super resonance condition ($^{13}$C = 9.54%; $^{15}$N = 10.89%; $^{18}$O = 6.6%). The final result was obtained when the average of 7x4 = 28 (take experiment of testing *E. coli* growth at 0.37-10% $^{15}$N as example) median values, and the corresponding standard error, were calculated. The p-values for non-terrestrial compositions were calculated using two-tailed, paired Student's t-test against the terrestrial composition (standard).

**Supplementary Figures**

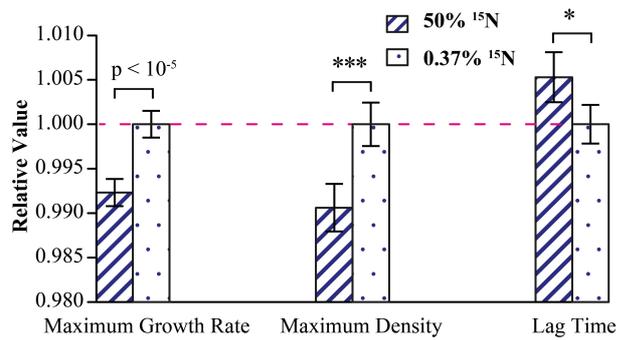

**Figure S1.** Growth parameters of *E. coli* at 50% of $^{15}$N in M9 minimal media, compared to normal isotopic conditions. * denotes p<0.05, ** denotes p<0.05, etc.

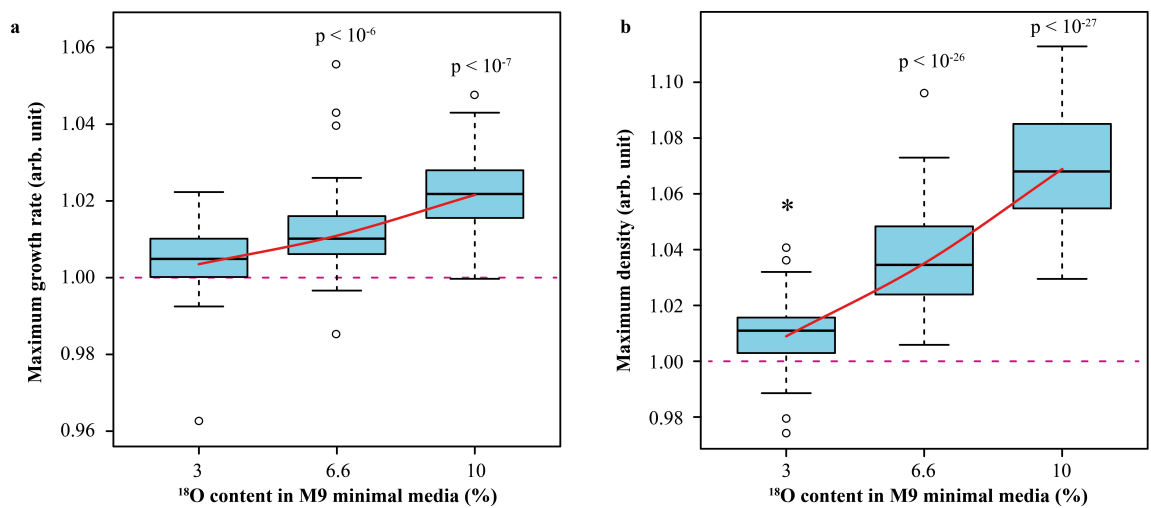

**Figure S2.** Maximum growth rate (a) and maximum density (b) of *E. coli* growth with $^{18}$O enriched in the water of M9 minimal media. * denotes p<0.05.

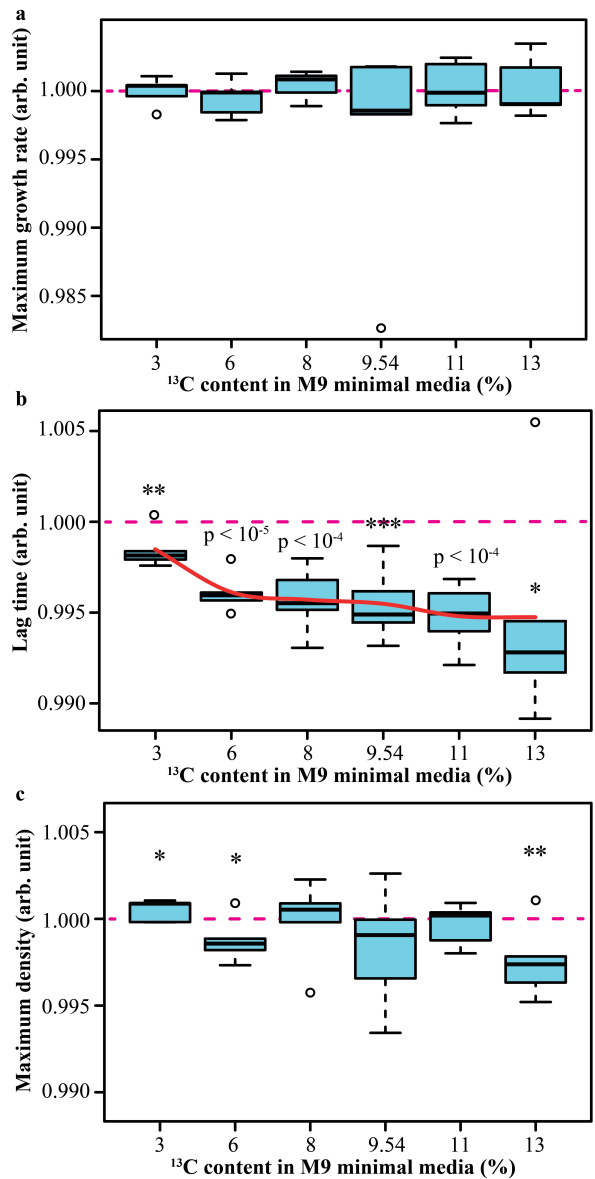

**Figure S3.** Growth parameters of *E. coli* growth with $^{13}$C enriched in the glucose of M9 minimal media. (a) Maximum growth rate, (b) lag time, (c) maximum density. * denotes p<0.05, ** denotes p<0.05, etc.

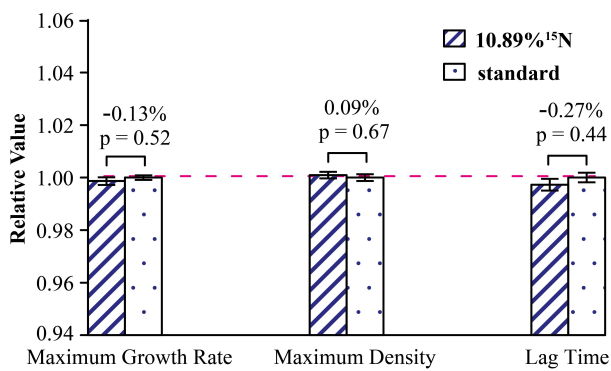

**Figure S4.** Growth parameters of *E. coli* growth with $^{15}$N enriched in the salt of M9 minimal media.

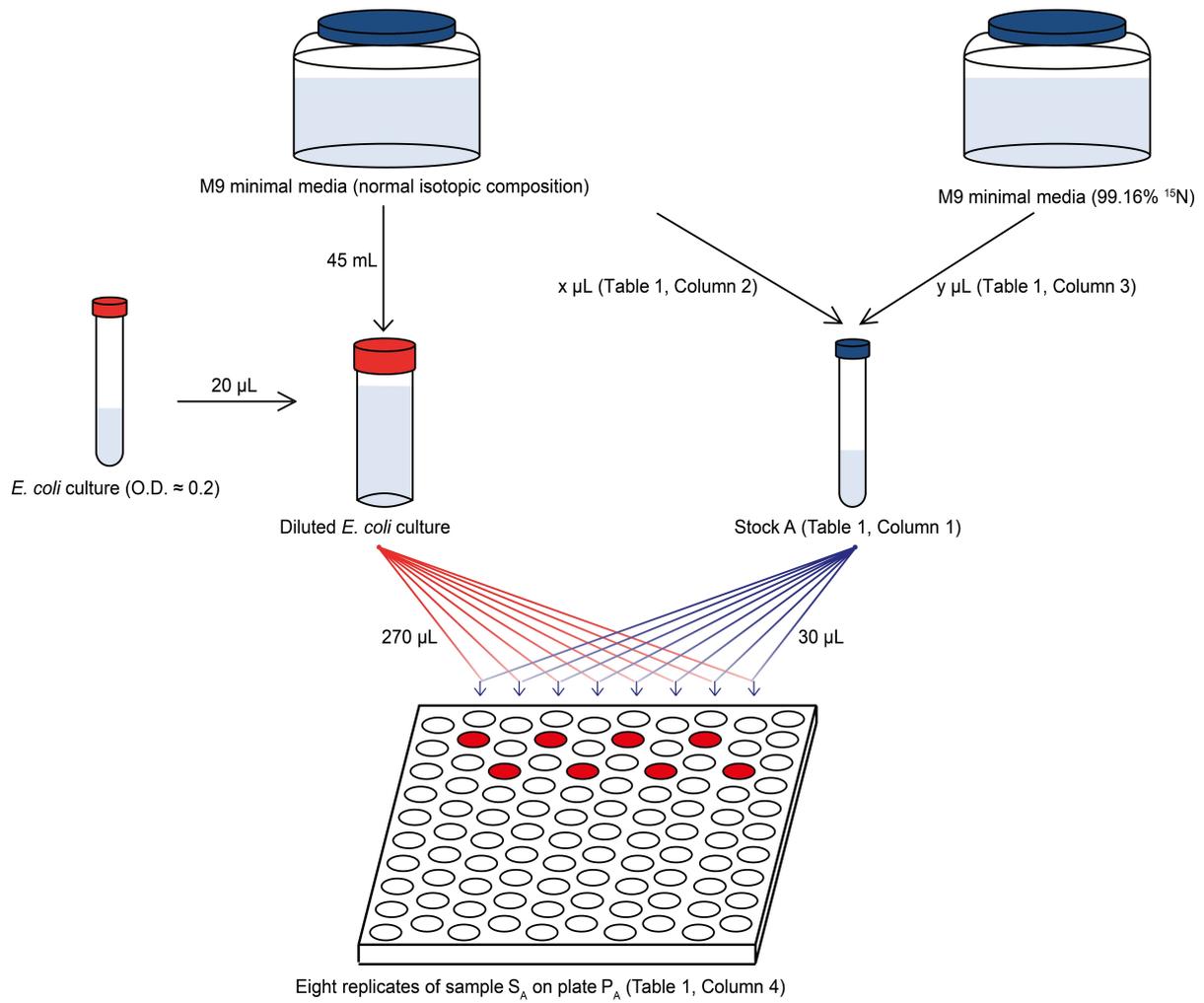

**Figure S5.** Workflow of sample preparation.

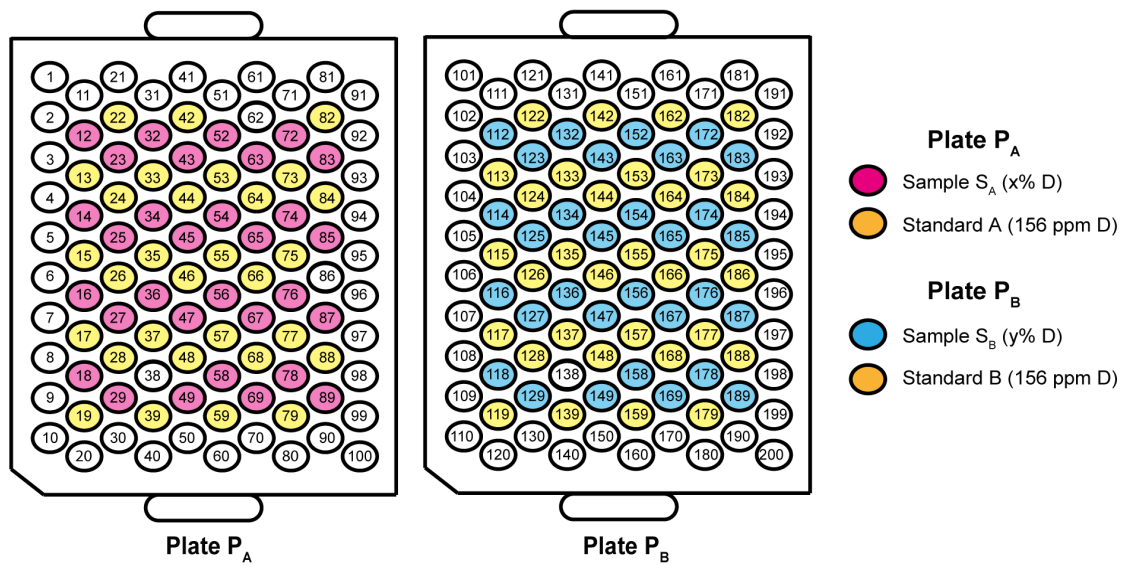

**Figure S6.** Sample configuration on the honeycomb well plates, adapted from reference 33.